\documentclass[aps,prb,twocolumn,groupedaddress,showpacs]{revtex4}
\usepackage{graphicx}

\begin{document}


\title{Temperature Effects on Exciton and Trion States in CdTe Quantum Well Structures}


\author{M. Fehr}
\affiliation{Technische Universit$\ddot{a}$t Munchen, D-85747
Garching} \affiliation{A. F. Ioffe Physico-Technical Institute,
194021 St. Petersburg, Russia}
\author{D. Andronikov, V. Kochereshko}
\affiliation{A. F. Ioffe Physico-Technical Institute, 194021 St.
Petersburg, Russia}
\author{S. A. Crooker}
\affiliation{National High Magnetic Field Laboratory, Los Alamos, New Mexico 87545 USA}
\author{G. Karczewski}
\affiliation{Institute of Physics, Polish Academy of Sciences, al.
Lotnikow 32/46, 02-668 Warsaw, Poland}

\date{\today}

\begin{abstract}
We study the temperature-dependent modifications of trion and
exciton photoluminescence (PL) spectra in modulation-doped
CdTe/CdMgTe quantum wells in high magnetic field.  We find that,
in magnetic field, the temperature-dependent redistribution of
exciton and trion PL intensities is opposite to that expected from
a simple Boltzmann distribution model. Solving a system of rate
equations that describe the exciton-trion energy levels, we
calculate the temperature dependence of the exciton and trion PL
intensities. The calculations show good agreement with the
experimental data.
\end{abstract}


\maketitle


\section{INTRODUCTION}
The many-electron problem is a central problem encountered in many
fields of modern physics, including plasma, nuclear and condensed
matter physics. In condensed matter systems, one of the most
important examples of a many-electron system is a multi-electron
complex in a semiconductor nanostructure. The primary of them is
the negatively charged "trion" - and exciton-electron complex
consisting of one hole bound to two electrons. The properties of
trions in semiconductors are in many respects similar to those of
the negatively charged hydrogen ion (one proton bound to two
electrons) observed in the spectra of stellar atmospheres. Trions
in semiconductor quantum wells were first experimentally observed
in 1993 [\onlinecite{kheng}]. Since then, trions have been
intensively studied in various semiconductor heterostructures.
Singlet and triplet trion states have been studied in
electron-doped quantum wells based on different semiconductor
materials and at different electron concentrations in magnetic
field. To date, however, many trion properties remain unresolved.
In this work we study temperature-dependent changes in the trion
and exciton photoluminescence spectra in modulation-doped
CdTe/CdMgTe quantum wells in high magnetic fields to 45 Tesla.

\section{Experiment and Results}

CdTe/Cd$_{0.7}$Mg$_{0.3}$Te structures with a single 100{\AA}
quantum well (QW) grown on (100) GaAs substrates are studied. An
iodine-doped $\delta$-layer is located 100{\AA} from the QW.  We
study a series of such heterostructures grown during one epitaxial
growth using a wedge-doping technique \cite{2}; the structures are
different only in the doping level in the $\delta$-layer (and
therefore the 2DEG density). The electron concentration in the QW
varies from 10$^{10}$ cm$^{ - 2}$ to 10$^{12}$ cm$^{ - 2}$.
Polarized photoluminescence (PL) from these samples was measured
with magnetic fields applied in the Faraday configuration. A
capacitor-driven 50T pulsed magnet having a 400 ms pulse duration
was used. The samples were excited by a frequency-doubled
diode-pumped YAG laser at $\lambda $=532 nm. A complete set of
field-dependent PL spectra was collected during each magnet pulse,
at a temperature of 1.6K, 4.2K and 15K. Optical fibers were used
for optical illumination of the sample and collection of PL, and
the PL was detected in both circular polarizations $\sigma^+ $ and
$\sigma^- $.

\begin{figure}
\includegraphics[scale=1]{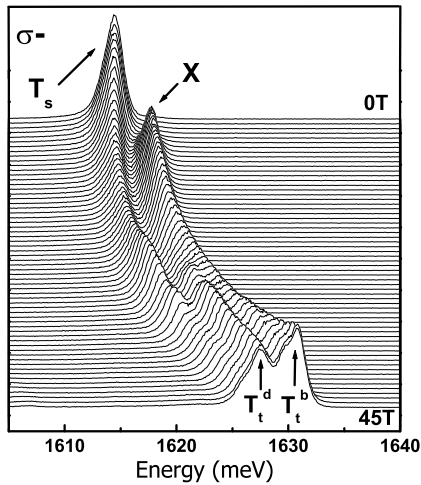}
\caption{\label{fig1}Set of PL spectra taken in magnetic fields
from 0 to 45 T from the sample with
$n_{e}=3\times10^{10}$~cm$^{-2}$ in $\sigma^-$ circular
polarization. $X$ indicates the exciton line, $T_{s}$ is the
singlet trion line, $T_{t}^{d}$ is the dark triplet trion line,
and $T_{t}^{b}$ is the bright triplet trion line.}
\end{figure}

Figure 1 shows a set of PL spectra taken from a
CdTe/Cd$_{0.7}$Mg$_{0.3}$Te quantum well structure having electron
concentration $n_{e}=3\times10^{10}$~cm$^{-2}$ in magnetic fields
from 0 to 45T at T=1.6K in $\sigma^- $ circular polarization. We
do not show here the spectra in $\sigma^+ $ circular polarization
because they are less informative.

In zero magnetic field the spectrum is dominated by the PL line of
the singlet trion state $T_{s}$. With increasing magnetic field
the recombination line of the exciton $X$ is revealed in the
spectra starting from $\sim$3T. In higher magnetic fields
(H$\geq$20T) an additional PL line from dark triplet trion states
$T_{t}^{d}$ emerges in the spectra. At the same time the singlet
trion line loses amplitude, while the exciton line retains its
intensity. In very high magnetic fields (H$\geq$35T) the PL line
of the bright triplet state $T_{t}^{b}$ separates from the exciton
line. The intensity of the bright triplet line is comparable to
that of the exciton. However, due to a small binding energy of
this state and appreciable widths of the lines, they partially
overlap. A detailed classification of the spectral lines of the
CdTe-based QW is given in \cite{3}.

\begin{figure}
\includegraphics[scale=0.45]{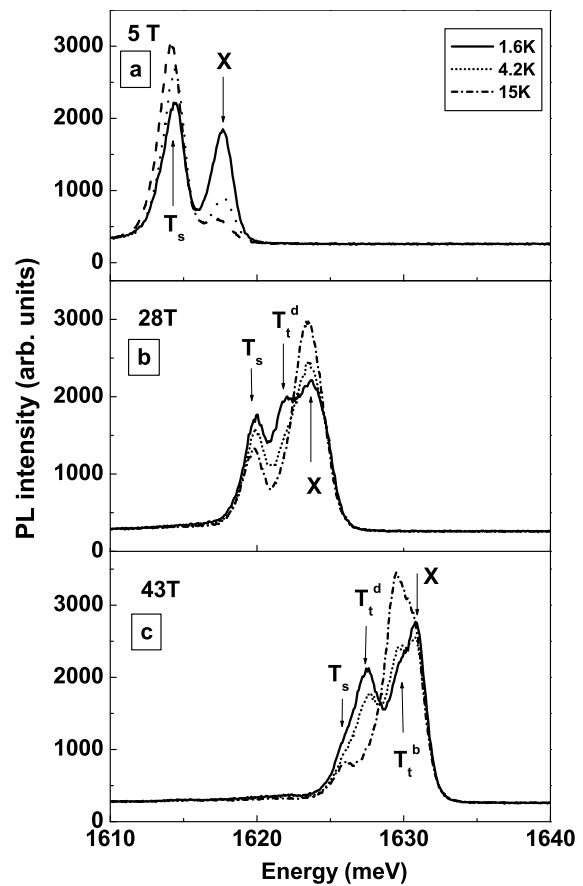}
\caption{\label{fig2}Photoluminescence spectra taken from the
sample with $n_{e}=3\times10^{10}$~cm$^{-2}$ in $\sigma^-$
circular polarization at three different magnitudes of the
magnetic field: 5T (a), 28T (b), and 43T (c) at 1.6K (solid line),
4.2K (dotted line), and 15K (dash-dotted line). The designation of
the PL lines is the same as in figure 1.}
\end{figure}

Fig. 2à shows temperature-induced modification of PL spectra taken
from this sample in   circular polarization in a magnetic field of
5T. It is evident that the singlet trion PL line ($T_{s}$) gains
intensity with increasing temperature, while the intensity of the
exciton PL line ($X$) falls . The integrated intensity is
conserved. Such redistribution of intensities is unusual, since
one should expect that increasing temperature should populate
higher energy states and deplete the lower-lying states according
to Boltzmann factor $exp(-\Delta E/KT)$  - in direct contrast to
the data. This observation directly indicates the nonequilibrium
nature of the observed PL.

An unusual temperature-dependent behavior of the trion and exciton
PL was also observed in high magnetic fields. Figure 2b shows the
PL spectra of the same sample at 28 Tesla for three temperatures:
1.6K, 4.2K and 15K. Along with the exciton ($X$) and the singlet
trion ($T_{s}$) PL lines, a PL line of the dark triplet trion
($T_{t}^{d}$) was observed in these spectra \cite{3, 4}. It is
obvious that in this case the temperature-induced redistribution
of the intensities of the different spectral lines also takes
place. However in this case the picture is reversed: the PL
intensities of singlet and dark triplet trion states fall when
temperatures rise, while the intensity of the higher-energy
exciton line increases. Since the temperature-dependent intensity
redistribution in low magnetic field (5T) is reversed from the one
observed in high field (28T), one may expect at some intermediate
field that no temperature-dependent intensity redistribution
between exciton and trion lines exists at all. Indeed, in the
range of the magnetic fields between 14 and 18T such
redistribution of intensities is not observed in the spectra. At
the highest magnetic fields the scenario changes yet again. Figure
2c shows temperature-dependent PL spectra at 43 Tesla. In magnetic
fields of this magnitude the PL intensities of the singlet trion
$T_{s}$ and the dark triplet trion $T_{t}^{d}$ fall with growing
temperature, while the intensity of the bright triplet PL line
$T_{t}^{b}$ significantly increases. The intensity of the exciton
line does not noticeably change. This also appears surprising
since the binding energy of the bright triplet trion $T_{t}^{b}$
($\sim$1meV at 40 Tesla) is small compared to the binding energy
of the dark triplet $T_{t}^{d}$ ($\sim$3.5meV at 40 Tesla).
Moreover, the binding energy of the $T_{t}^{b}$ state is
comparable to temperature.

\section{Discussion}

This unusual temperature-dependent behavior of exciton and trion
PL indicates its non-equilibrium nature. The observed phenomena
are evidently caused by the influence of temperature on the
kinetics of the final state populations, rather than by the
Boltzmann factor. Indeed, as the scheme of trion formation shows
in Figure 3, a singlet trion can only be formed when an electron
from the lower Zeeman sublevel and an exciton from one of the
higher sublevels bind, or when an electron from an upper sublevel
binds with an exciton from one of the lower sublevels. In high
enough magnetic fields, 5T in our sample, all the background
electrons are fully spin-polarized, i.e. they are all in the
lowest Zeeman sublevel. At the same time, since the excitons are
excited by high-energy photons, the upper exciton Zeeman
components can be partially populated even at low temperatures. As
a result, the lower trion Zeeman component (- 3/2) which is active
in $\sigma^-$ polarization will be populated by an electron (+1/2)
bound with a dark exciton (-2), while the upper trion Zeeman
component, active in $\sigma^+$ polarization will be populated by
an electron (+1/2) bound with a bright exciton (+1).

\begin{figure}
\includegraphics[scale=0.9]{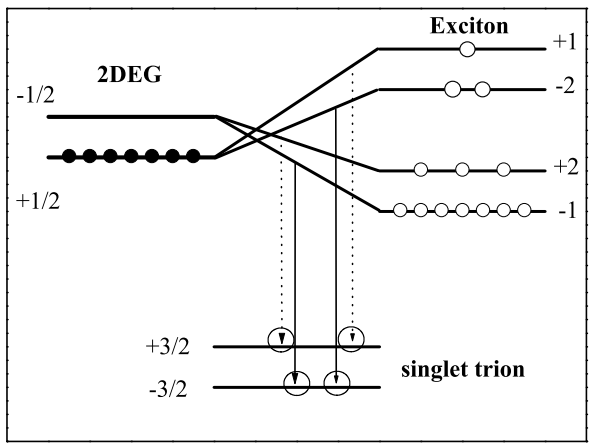}
\caption{\label{fig3}The scheme of the singlet trion formation
mechanism in the magnetic field. Electron, exciton, and singlet
trion states with different spins are shown. $\pm1$ correspond to
bright exciton states, $\pm2$ correspond to dark excitons.}
\end{figure}

When the temperature increases, a redistribution of electron and
exciton populations within these sublevels occurs according to the
Boltzmann factor. The redistribution will change the sublevel
populations of dark (+2) and bright (-1) excitons in favor of the
dark exciton and background electrons (+1/2) and (-1/2) because of
the small value of the hole g-factor \cite{5}. As a result, the
exciton PL intensity in   polarization will fall, while the
intensity of trion radiation in the same polarization will rise.
As seen from Figure 3, the temperature-induced redistribution of
exciton and trion PL intensities in $\sigma^+ $ polarization
should be opposite and should comply with a Boltzmann
distribution, as also observed in the experiment
\cite{Polonica04}. These conclusions are confirmed by calculating
exciton and trion PL intensities. In order to calculate the PL
intensities, we solved a system of rate equations with parameters
used in \cite{3}.

\begin{equation}
\label{eq3} \frac{\partial n_{i}}{\partial t}=\sum( n_{j} w_{ji} -
n_{i}w_{ij}) + g_{i} - n_{i}/\tau^{0}_{i}.
\end{equation}

Here $n_{i}$ is the population of the sublevel, $w_{ij}$ is the
transition rate from $i$-th sublevel to $j$-th sublevel, $g_{i}$
is the generation rate on $i$-th sublevel, $\tau^{0}_{i}$ is the
radiative lifetime of the $i$-th sublevel. Figure 4a shows the
calculated results of the temperature dependences of the exciton
and trion PL intensities and the experimental data for 1.6K, 4.2K
and 15K temperatures at 5T. The experimental data and calculated
curves show very good agreement. We have also performed a
calculation for other values of magnetic field. As seen from the
experiment, in high magnetic fields the temperature dependence of
the spectra is qualitatively different from the one in low fields.
Figures 4b and 4c present the calculation results of temperature
dependences of exciton and trion PL for 28T and 43T magnetic
fields. The qualitative agreement between experiment and
calculation is also observed here. According to theoretical
predictions \cite{6, 7} and experimental observations \cite{3, 8},
in high magnetic fields triplet trion states appear along with the
singlet states. The dark triplet line $T_{t}^{d}$ is distinctly
observed in the PL spectra starting from 15T, while the bright
triplet line $T_{t}^{b}$ appears as separate spectral line at
fields exceeding 35T. At low fields this line merges with the
exciton PL line. This explains the intensity increase of the
"exciton PL" in the field range between 22T and 30T. At higher
field, when the exciton and bright triplet lines are well
separated, it is evident that increasing temperature leads to the
enhancement the bright triplet line instead of the exciton line.
Therefore, figures 4b and 4c also show the calculated curve for
the sum of exciton and bright triplet intensities.

\begin{figure}
\includegraphics[scale=0.4]{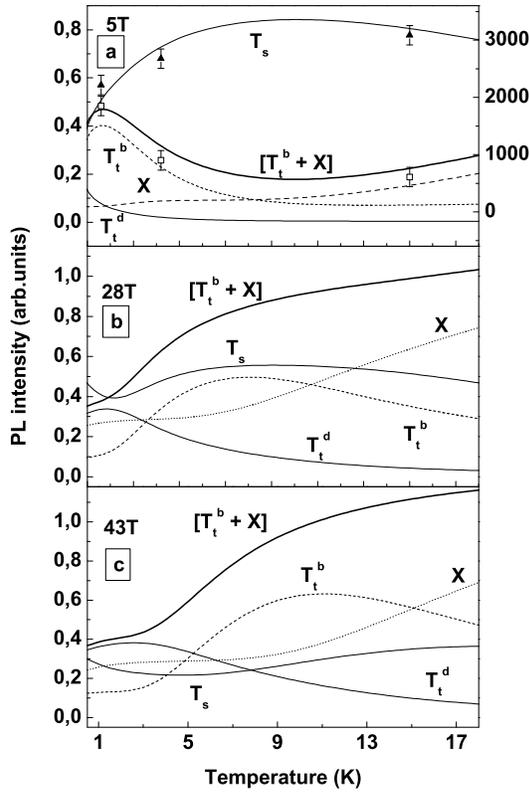}
\caption{\label{fig4} Calculation results of temperature
dependences of exciton and trion PL for 5T (a), 28T(b), and 43T(c)
magnitudes of the magnetic fields. The designations of the PL
features are the same as in previous figures. $T_{t}^{b}$+$X$
stands for total contribution from the bright triplet and exciton
PL lines. In (a) closed triangles and open rectangles stand for
scaled experimental values of singlet trion and exciton PL
intensities correspondingly. }
\end{figure}

We explain the enhancement of the bright triplet trion intensity
by the influx into this state from the reservoir of dark triplet
trions. The concentration of the dark triplet states is very high
due to its short formation time and its long lifetime, which is at
least three orders of magnitude different. This qualitative
argument is fully supported by the calculated PL intensities of
all the spectral lines, presented in figures 4b and 4c.

\section{Conclusions}

Modifications of trion and exciton PL spectra with temperature in
modulation-doped CdTe/CdMgTe quantum wells in magnetic fields have
been studied. We found that in magnetic field the temperature
redistribution of the exciton and trion PL line intensities is
opposite to the one expected in a simple Boltzmann distribution
model. Solving a system of kinetic equations that describe the
exciton-trion system, we calculate the temperature dependence of
the exciton and trion PL intensities. The calculated dependences
show good agreement with the experimental data.

\begin{acknowledgments}
This work was supported in part by RFBR and Presidium RAS grants.
M.F. is very grateful to DAAD foundation.
\end{acknowledgments}

\bibliography{fehr}

\end{document}